\documentclass[twocolumn]{aastex63}
\usepackage{xcolor}
\usepackage{amsmath,amstext}
\usepackage{graphicx}
\usepackage{subfigure}
\usepackage{tabularx}
\usepackage{enumitem}
\usepackage{appendix}

\def\sn{SN\,2023ixf}

\received{TBD}
\revised{TBD}
\accepted{TBD}
\published{TBD}
\submitjournal{ApJ}

\shorttitle{Millimeter Observations of \sn}
\shortauthors{Berger et al.}

\begin{document}

\title{\bf \Large Millimeter Observations of the Type II SN\,2023ixf: Constraints on the Proximate Circumstellar Medium}

\author[0000-0002-9392-9681]{Edo~Berger}
\affiliation{Center for Astrophysics \textbar{} Harvard \& Smithsonian, 60 Garden Street, Cambridge, MA 02138-1516, USA}
\correspondingauthor{Edo Berger}
\email{eberger@cfa.harvard.edu}

\author[0000-0002-3490-146X]{Garrett~K.~Keating}
\affiliation{Center for Astrophysics \textbar{} Harvard \& Smithsonian, 60 Garden Street, Cambridge, MA 02138-1516, USA}

\author[0000-0003-4768-7586]{Raffaella Margutti}
\affiliation{Department of Astronomy, University of California, Berkeley, CA, 94720}
\affiliation{Department of Physics, University of California, Berkeley, CA, 94720}

\author[0000-0003-2611-7269]{Keiichi Maeda}
\affiliation{Department of Astronomy, Kyoto University, Kitashirakawa-Oiwake-cho, Sakyo-ku, Kyoto 606-8502, Japan}

\author[0000-0002-8297-2473]{Kate D.~Alexander}
\affiliation{Department of Astronomy/Steward Observatory, 933 North Cherry Avenue, Rm. N204, Tucson, AZ 85721-0065, USA}

\author[0000-0001-7007-6295]{Yvette Cendes}
\affiliation{Center for Astrophysics \textbar{} Harvard \& Smithsonian, 60 Garden Street, Cambridge, MA 02138-1516, USA}

\author[0000-0003-0307-9984]{Tarraneh Eftekhari}
\altaffiliation{NASA Einstein Fellow}
\affiliation{Center for Interdisciplinary Exploration and Research in Astrophysics (CIERA) and Department of Physics and Astronomy, Northwestern University, Evanston, IL 60208, USA}

\author[0000-0003-0685-3621]{Mark Gurwell}
\affiliation{Center for Astrophysics \textbar{} Harvard \& Smithsonian, 60 Garden Street, Cambridge, MA 02138-1516, USA}

\author[0000-0002-1125-9187]{Daichi Hiramatsu}
\affiliation{Center for Astrophysics \textbar{} Harvard \& Smithsonian, 60 Garden Street, Cambridge, MA 02138-1516, USA}

\author[0000-0002-9017-3567]{Anna Y. Q.~Ho}
\affiliation{Department of Astronomy, Cornell University, Ithaca, NY 14853, USA}

\author[0000-0003-1792-2338]{Tanmoy Laskar}
\affiliation{Department of Physics \& Astronomy, University of Utah, Salt Lake City, UT 84112, USA}
\affiliation{Department of Astrophysics/IMAPP, Radboud University, P.O. Box 9010, 6500 GL, Nijmegen, The Netherlands}

\author[0000-0002-1407-7944]{Ramprasad~Rao}
\affiliation{Center for Astrophysics \textbar{} Harvard \& Smithsonian, 60 Garden Street, Cambridge, MA 02138-1516, USA}

\author[0000-0003-3734-3587]{Peter K.\ G.\ Williams}
\affiliation{Center for Astrophysics \textbar{} Harvard \& Smithsonian, 60 Garden Street, Cambridge, MA 02138-1516, USA}

\begin{abstract}
We present 1.3 mm (230 GHz) observations of the recent and nearby Type II supernova, \sn, obtained with the Submillimeter Array (SMA) at $2.6-18.6$ days after explosion. The observations were obtained as part the SMA Large Program {\it POETS} (Pursuit of Extragalactic Transients with the SMA).  We do not detect any emission at the location of \sn, with the deepest limits of $L_\nu(230\,{\rm GHz})\lesssim 8.6\times 10^{25}$ erg s$^{-1}$ Hz$^{-1}$ at 2.7 and 7.7 days, and $L_\nu(230\,{\rm GHz})\lesssim 3.4\times 10^{25}$ erg s$^{-1}$ Hz$^{-1}$ at 18.6 days.  These limits are about a factor of 2 times dimmer than the mm emission from SN\,2011dh (IIb), about an order of magnitude dimmer compared to SN\,1993J (IIb) and SN\,2018ivc (IIL), and about 30 times dimmer than the most luminous non-relativistic SNe in the mm-band (Type IIb/Ib/Ic). Using these limits in the context of analytical models that include synchrotron self-absorption and free-free absorption we place constraints on the proximate circumstellar medium around the progenitor star, to a scale of $\sim 2\times 10^{15}$ cm, excluding the range $\dot{M}\sim {\rm few}\times 10^{-6}-10^{-2}$ M$_\odot$ yr$^{-1}$ (for a wind velocity, $v_w=115$ km s$^{-1}$, and ejecta velocity, $v_{\rm eje}\sim (1-2)\times 10^4$ km s$^{-1}$). These results are consistent with an inference of the mass loss rate based on optical spectroscopy ($\sim 2\times 10^{-2}$ M$_\odot$ yr$^{-1}$ for $v_w=115$ km s$^{-1}$), but are in tension with the inference from hard X-rays ($\sim 7\times 10^{-4}$ M$_\odot$ yr$^{-1}$ for $v_w=115$ km s$^{-1}$). This tension may be alleviated by a non-homogeneous and confined CSM, consistent with results from high-resolution optical spectroscopy.
\end{abstract}

\keywords{
\href{https://vocabs.ardc.edu.au/repository/api/lda/aas/the-unified-astronomy-thesaurus/current/resource.html?uri=http://astrothesaurus.org/uat/1668}{Supernovae (1668)}; 
\href{https://vocabs.ardc.edu.au/repository/api/lda/aas/the-unified-astronomy-thesaurus/current/resource.html?uri=http://astrothesaurus.org/uat/304}{Core-collapse supernovae (304)}; 
\href{https://vocabs.ardc.edu.au/repository/api/lda/aas/the-unified-astronomy-thesaurus/current/resource.html?uri=http://astrothesaurus.org/uat/1731}{Type II supernovae (1731)}; 
\href{https://vocabs.ardc.edu.au/repository/api/lda/aas/the-unified-astronomy-thesaurus/current/resource.html?uri=http://astrothesaurus.org/uat/732}{Massive stars (732)};
\href{https://vocabs.ardc.edu.au/repository/api/lda/aas/the-unified-astronomy-thesaurus/current/resource.html?uri=http://astrothesaurus.org/uat/1613}{Stellar mass loss (1613)}
\href{https://vocabs.ardc.edu.au/repository/api/lda/aas/the-unified-astronomy-thesaurus/current/resource.html?uri=http://astrothesaurus.org/uat/241}{Circumstellar matter (241)}
}

\section{Introduction} 
\label{sec:intro}

Over the past few decades observations of core-collapse supernovae (CCSNe) ranging from radio to X-rays have revealed that massive stars may undergo enhanced or eruptive mass loss in the final years and decades prior to core-collapse (e.g., \citealt{Smith2014}). The resulting circumstellar medium (CSM) around the progenitor stars can be probed using optical spectroscopy shortly after explosion, which can reveal emission from the ionized CSM  (so-called flash spectroscopy; e.g., \citealt{Gal-yam2014,Yaron2017}); optical/UV observations of the rising light curves, which can reveal energy deposition from CSM-SN ejecta shock interaction (e.g., \citealt{NakarPiro2014,Moriya2018,Hiramatsu2021}); X-ray observations, which track thermal bremsstrahlung emission from CSM-SN ejecta shock interaction and absorption by neutral CSM; and radio/millimeter observations which trace synchrotron emission due to the acceleration of relativistic electrons in CSM-SN shock interaction (e.g., \citealt{Chevalier17}).

\sn\ was discovered on 2023 May 19.727 UT shortly after explosion in M101 \citep{Itagaki23}, at $d\approx 6.9$ Mpc \citep{Riess2022}, and was classified as a Type II SN shortly thereafter \citep{Perley23}. Thanks to its early discovery and proximity, extensive data across the electromagnetic spectrum is being collected, including early optical spectra that reveal flash ionization features \citep{jg23,Smith23} and narrow absorption due to the pre-shocked CSM \citep{Smith23}; optical light curves, which point to early excess emission \citep{Hosseinzadeh23}; and hard X-ray observations, indicating a large neutral hydrogen column density in the vicinity of the SN \citep{Grefenstette23}. Pre-explosion {\it Hubble Space Telescope}, {\it Spitzer Space Telescope}, and ground-based observations point to a variable and dusty red supergiant progenitor \citep{Kilpatrick23,Neustadt23,Pledger23}. Here, we report on early observations obtained with the SMA spanning about 2.7 to 18.6 days after the estimated time of first light, and use these data to constrain the CSM density around the progenitor star.  We present the data in \S\ref{sec:obs} and compare to previous mm-band data for CCSNe in \S\ref{sec:comp} and to numerical and analytical models in \S\ref{sec:models}. We discuss and summarize the results in \S\ref{sec:conc}.

\section{Observation and Data Analysis} 
\label{sec:obs}

Following the discovery and classification of \sn, we used the SMA on several occasions to observe the SN, starting on 2023 May 21.17, about 2.4 days after the estimated time of first light (2023 May 18.74 UT; e.g., \citealt{Hosseinzadeh23}; Hiramatsu et al.~in prep.) The observations were obtained as part of the new Large Project {\it POETS} (Pursuit of Extragalactic Transients with the SMA; project 2022B-S046, PI: Berger). The observations are summarized in Table~\ref{tab:sma}. During these observations, the SMA was tuned to an LO frequency of 225.5 GHz, providing spectral coverage between 209.5--221.5 and 229.5--241.5 GHz. Across all nights, 3C454.3 was observed as a bandpass calibrator, Ceres was observed as a flux calibrator, and J1419+543 and J1506+426 were observed as gain calibrators, with a 15-minute cycle time cadence. 

\begin{figure}[t]
    \centering
    \includegraphics[width=0.45\textwidth]{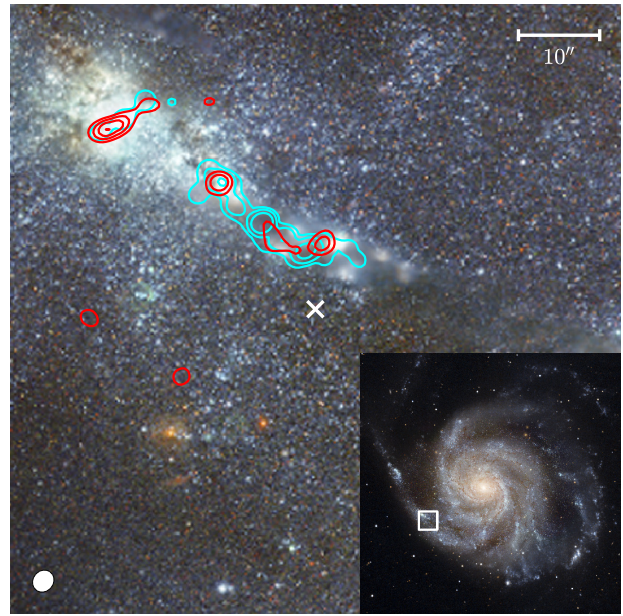}
\caption{Map from the aggregate SMA data from all 6 epochs of observations, of the region around \sn\ (white cross) overlaid on a false-color \textit{Hubble Space Telescope} image, with the synthesized beam shown in the lower-left (white ellipse). We show contours of the 230 GHz (1.3 mm) continuum emission in red, and the CO(2-1) line emission in cyan. Continuum contours correspond to 0.6, 0.9, 1.2, and 1.5 mJy; CO(2-1) contours correspond to 2, 4, 8 Jy\, km\, s$^{-1}$ (uncorrected for primary beam effects). No continuum emission is detected at the position of \sn, which is located near NGC\,5461, an \ion{H}{2} region at the southeastern edge of M101 (white box in inset).}
    \label{fig:image}
\end{figure}

\begin{deluxetable}{lccc}
\tablecolumns{4}
\tablecaption{SMA Observations of \sn}
\tablehead{
UT Date     &
$\delta t$  &
$F_\nu\,^a$ &
$L_\nu\,^a$ \\
            &
(d)         &
(mJy)       &
($10^{26}$ erg s$^{-1}$ Hz$^{-1}$)
}
\startdata
May 21.17--21.63 & 2.66 & $<1.5$ & $<0.86$ \\
May 22.17--22.29 & 3.49 & $<3.3$ & $<1.89$ \\
May 24.51--24.62 & 5.83 & $<4.2$ & $<2.40$ \\
May 26.34--26.62 & 7.74 & $<1.5$ & $<0.86$ \\
May 28.44--28.62 & 9.79 & $<2.4$ & $<1.37$ \\
June 6.09--6.60 & 18.60 & $<0.6$ & $<0.34$ \\
\enddata
\tablecomments{Phases ($\delta t$) are given relative to an estimated time of first light of 2023 May 18.74 UT and are at the mid-point of each observation.\\
$^a$ Limits are 3 times the image rms.}
\label{tab:sma}
\end{deluxetable}

Analysis of the data was performed using the SMA COMPASS pipeline (Keating et al., in prep.), which flags spectral data based on outliers in amplitude when coherently averaging over increasing time intervals for each channel within each baseline, as well as baselines where little-to-no coherence is seen on calibrator targets. Flux calibration was performed using the Butler-JPL-Horizons 2012 \citep{2012ALMAM.594....1B} model for Ceres. The data were imaged, and deconvolution was performed via the CLEAN algorithm \citep{1974A&AS...15..417H}.  

We do not detect emission from the location of \sn\ in any of our observations, with root-mean-square noise levels spanning about $0.2-1.4$ mJy. The corresponding luminosity limits are $\lesssim 3.4\times 10^{25}-2.4\times 10^{26}$ erg s$^{-1}$ Hz$^{-1}$ (Table~\ref{tab:sma}).  An SMA continuum and CO(2-1) line map are shown in Figure~\ref{fig:image} overlaid on a {\it Hubble Space Telescope} image of the field.

\begin{figure}[t]
    \centering
    \includegraphics[width=0.51\textwidth]{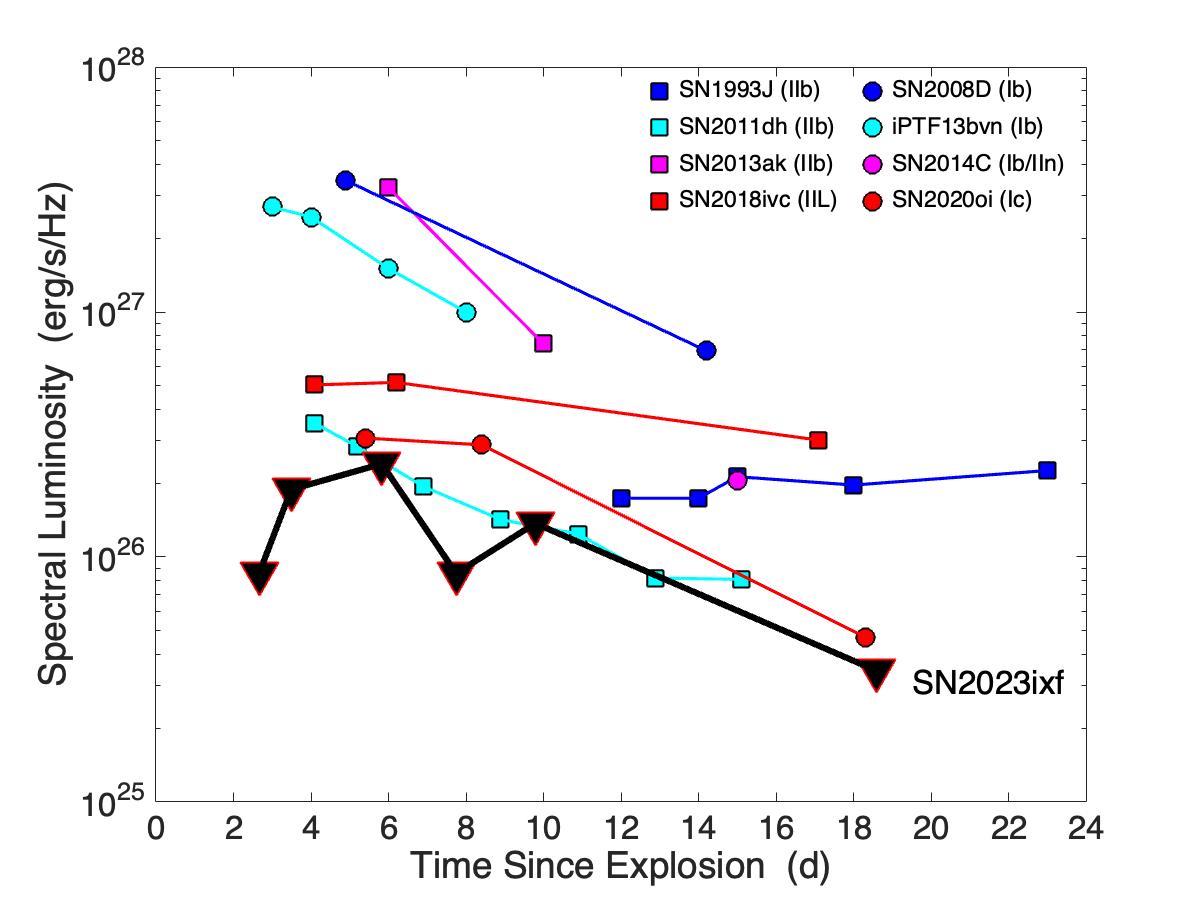}
\caption{SMA 230 GHz upper limits for \sn\ (black triangles; $3\sigma$) compared to the existing sample of CCSNe with early mm-band detections (squares: Type II; circles: Type Ib/c; \citealt{Weiler2007,Soderberg2008,Gorosabel2010,Horesh2013_iptf,Horesh2013,sn2013ak,sn2013ak_2,Zauderer2014,Maeda2021}).  
}
    \label{fig:lc}
\end{figure}

\section{Comparison to Other Core-Collapse Supernovae}
\label{sec:comp}

In Figure~\ref{fig:lc} we show the SMA spectral luminosity upper limits for \sn\ in comparison to the mm-band ($100-250$ GHz) light curves of several nearby CCSNe (Type II and Ib/c) on timescales of $\sim 3-23$ days.  Previously detected SNe span luminosities from only a factor of about 2 times higher than our limits (e.g., SN\,2011dh; \citealt{horesh13}) and up to $\sim 3\times 10^{27}$ erg s$^{-1}$ Hz$^{-1}$, more than an order of magnitude more luminous than our limits for \sn\ (e.g., SN\,2018ivc; \citealt{Maeda23}).  In recent work \citet{Maeda23} inferred mass loss rates (scaled to a wind velocity of 100 km s$^{-1}$) of $\sim 10^{-3}$ M$_\odot$ yr$^{-1}$ for SN\,2018ivc, $\sim 10^{-4}$ M$_\odot$ yr$^{-1}$ for SN\,1993J, and $\sim 10^{-5}$ M$_\odot$ yr$^{-1}$ for SN\,2011dh based on mm-band data. Comparable mass loss rates, $\sim {\rm few}\times 10^{-6}-10^{-4}$ M$_\odot$ yr$^{-1}$ have been inferred for other Type IIb SNe based on cm-band data (e.g., \citealt{Nayana22} and references therein). We note that all these comparison SNe are of Type IIb or IIL withmore stripped and compact progenitors than that of \sn\ \citep{Kilpatrick23,Neustadt23} and likely have higher velocity ejecta that boost their mm-band (and radio) synchrotron emission.

\section{CSM Interaction Models} 
\label{sec:models}

We use the non-detections at 230 GHz to place constraints on the CSM around the progenitor of \sn\ on a scale of $\sim 2\times 10^{14}-2\times 10^{15}$ cm (for an assumed ejecta velocity of $\sim 10^4$ km s$^{-1}$).  In the mm band the expected emission is due to synchrotron radiation arising from shock interaction of the SN ejecta with the CSM. We consider the effects of both synchrotron self-absorption and free-free absorption by an external CSM, following the well established models of \citet{Weiler86} and \citet{Chevalier98}, which have been used to interpret the radio/mm emission from previous CCSNe.  

In the high-density limit, free-free absorption is expected to dominate, with the optical depth given by \citet{Weiler86}: $\tau_{\rm ff}=K_2(\nu/5\,{\rm GHz})^{-2.1}t_d^{-3}$, where $t_d$ is time in days since explosion, and we have assumed that the ejecta are roughly in free expansion (leading to the temporal power law index of $-3$). Setting $\tau_{\rm ff}(230\,{\rm GHz})\gtrsim 1$ at 2.7 and 18.6 days we find $K_2\gtrsim 6.1\times 10^4$ and $\gtrsim 2\times 10^7$, respectively. The mass loss rate is given by \citep{Weiler86}:
\begin{multline}
    \dot{M}\approx 10^{-8}\,M_\odot\, {\rm yr}^{-1}\,
    \tau_{\rm ff,5\,{\rm GHz}}^{0.5}\,
    t_d^{1.5}\,
    (v_{\rm ej}/10^4\,{\rm km\,s^{-1}})^{1.5}\, \\
    \times (T_e/10^4\,{\rm K})^{0.68}\,
    (v_w/10\,{\rm km\,s^{-1}}),
\end{multline}
where $T_e$ is the electron temperature, and $v_w$ is the CSM wind velocity. Using our SMA limits at 2.7 and 18.6 days, with $v_w=115$ km s$^{-1}$ \citep{Smith23}, $v_{\rm ej}=10^4$ km s$^{-1}$, and $T_e=10^4$ K we find $\dot{M}\gtrsim 10^{-4}$ and $\gtrsim 4\times 10^{-2}$ M$_\odot$ yr$^{-1}$, respectively.

In the case of synchrotron self-absorption only, the inferred mass loss rate is much lower, and is given by \citep{Chevalier98}:
\begin{multline}
\dot{M}\approx 8.2\times 10^{-7}\,M_\odot\, {\rm yr}^{-1}\,
\epsilon_{B,-1}^{-1}\,
(\epsilon_e/\epsilon_B)^{-8/19}\,
L_{26}^{-4/19}\, \\
\times \nu_{230\,{\rm GHz}}^2\,
t_d^2\,
(v_w/10\,{\rm km\,s^{-1}}),
\end{multline}
where $\epsilon_e$ and $\epsilon_B$ are the post-shock energy fractions in the relativistic electrons and magnetic fields, respectively.  Using our limits at 2.7 and 18.6 days, and assuming $\epsilon_e=\epsilon_B=0.1$, we find $\dot{M}\lesssim 7\times 10^{-5}$ and $\lesssim 4\times 10^{-3}$ M$_\odot$ yr$^{-1}$, respectively.

\begin{figure}[t]
    \centering
    \includegraphics[width=0.475\textwidth]{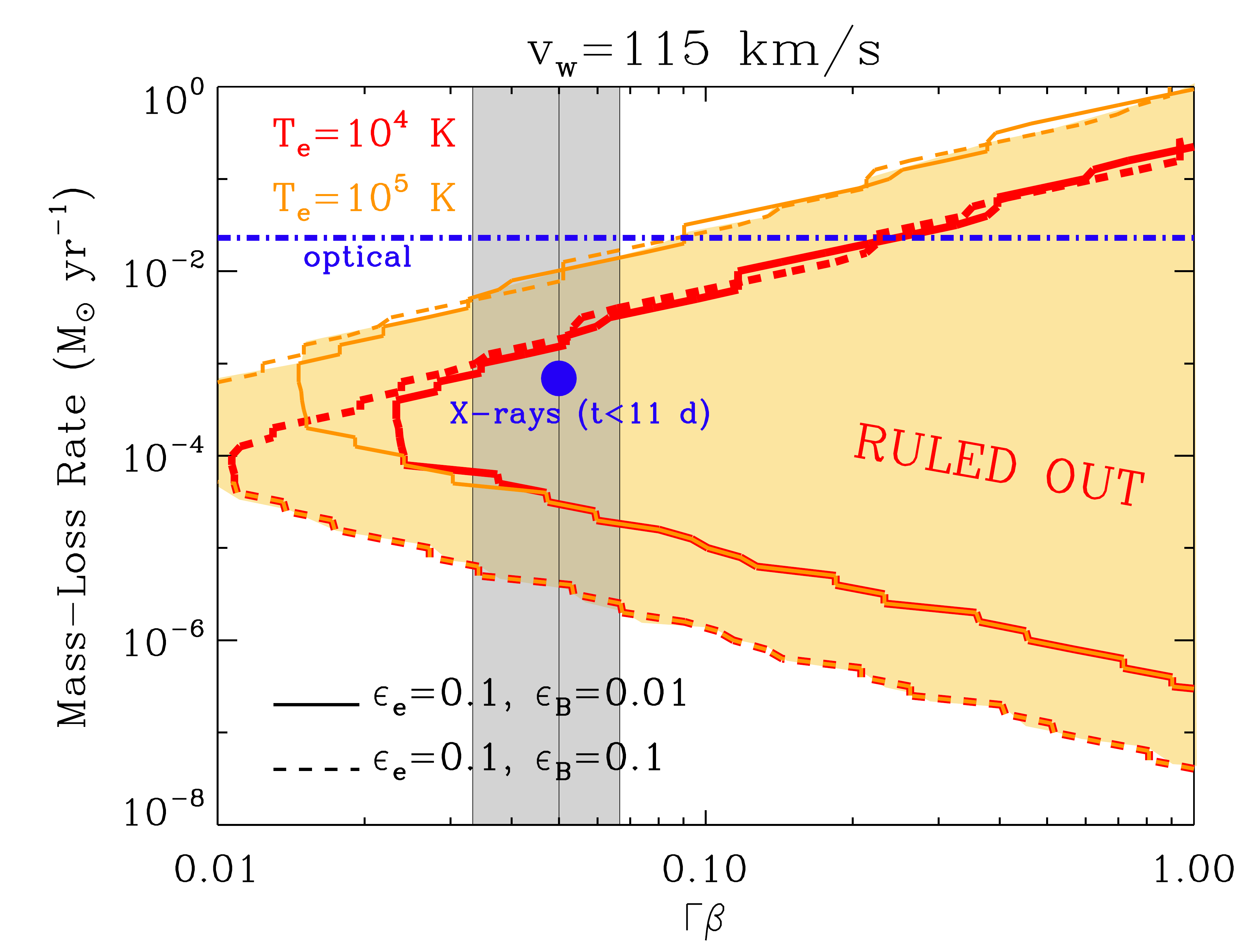}
\caption{The phase-space of progenitor mass-loss rate versus ejecta velocity that is {\it excluded} by our most constraining SMA limits at 2.7, 7.7, and 18.6 days. The models \citep{Chevalier98} assume $v_w=115$ km s$^{-1}$ \citep{Smith23} and are shown for two sets of equipartition parameters ($\epsilon_e$ and $\epsilon_B$), an electron power law distribution with $N_e(\gamma_e)\propto \gamma_e^{-p}$ with $p=3$, a volume filling factor of 0.5, and two values for the electron temperature ($T_e$).  The vertical grey shaded region indicates an ejecta velocity range of $v_{\rm ej}=(1-2)\times 10^4$ km s$^{-1}$. The blue dot marks the value of $\dot M$ inferred from X-ray observations acquired at $\lesssim 11$ days \citep{Grefenstette23}, while the horizontal blue dot-dashed line marks the value of $\dot M$ inferred from optical spectra at $\lesssim 14$ days \citep{jg23}. Both values have been updated to $v_w=115$ km s$^{-1}$.}
    \label{fig:mdot}
\end{figure}

A full calculation of the effects of free-free absorption and synchrotron self-absorption (following \citealt{Chevalier98}) is shown in Figure~\ref{fig:mdot}, where we constrain the joint phase-space of $\dot{M}$ and $v_{\rm ej}$ ruled out by our SMA data. In this calculation we determine the mm-band luminosity for a range of ejecta velocities and mass loss rates using an input kinetic energy, $v_w=115$ km s$^{-1}$, and assuming two sets of equipartition values ($\epsilon_e=\epsilon_B=0.1$ and $\epsilon_e=0.1, \epsilon_B=0.01$) and two values for $T_e$ ($10^4$ and $10^5$ K). We find that for a reasonable range of $v_{\rm ej}=(1-2)\times 10^4$ km s$^{-1}$ we can {\it exclude} a mass loss rate range of $\dot{M}\sim {\rm few}\times 10^{-6}-10^{-2}$ M$_\odot$ yr$^{-1}$ (for $\epsilon_e=\epsilon_B=0.1$ and $T_e=10^5$ K).

\begin{figure*}[t]
    \centering
    \includegraphics[width=0.45\textwidth]{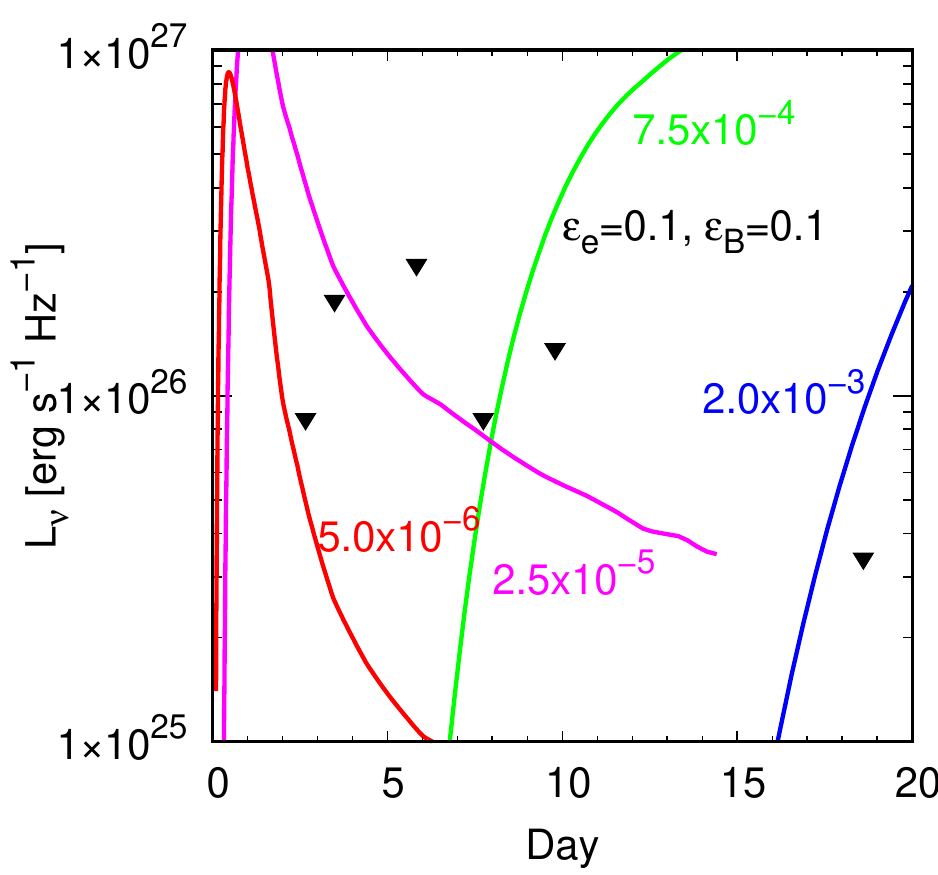}
     \includegraphics[width=0.45\textwidth]{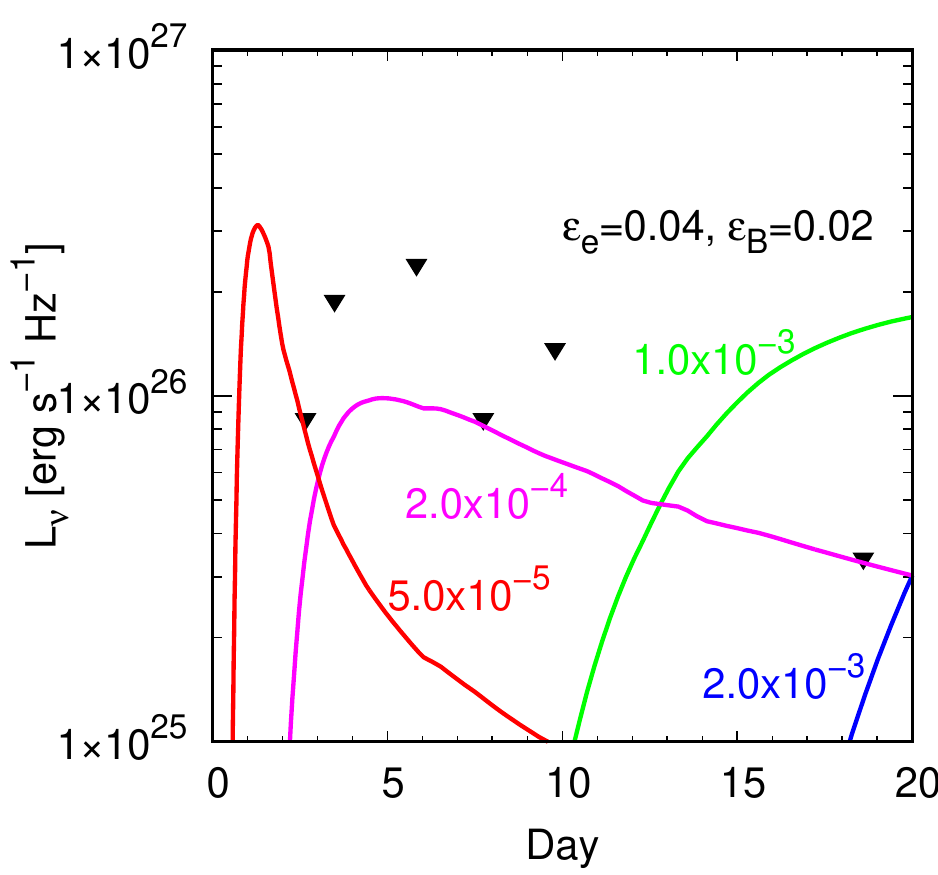}
\caption{Model light curves at 230 GHz (lines) using the formalism of \citet{Maeda2021} and \citet{Maeda23} for several choices of the mass loss rate and for two sets of equipartition parameters: $\epsilon_e=\epsilon_B=0.1$ ({\it Left}) and $\epsilon_{\rm e} = 0.04$ and $\epsilon_{\rm B} = 0.02$ ({\it Right}; calibrated with models for SN\,2020oi: \citealt{Maeda23}).  The triangles are the upper limits for \sn.}
    \label{fig:lcmodel}
\end{figure*}

The constraints shown in Figure~\ref{fig:mdot} are demonstrated by a comparison between model light curves and the observed upper limits in Figure~\ref{fig:lcmodel}. Here, we use the model prescriptions described in \citet{Maeda2021} and \citet{Maeda23}. We adopt a CSM with a wind profile truncated at $2\times 10^{15}$ cm to represent a ``confined'' CSM. The ejecta mass and energy are set to be $11 M_\odot$ and $10^{51}$ erg, and free-free absorption is included assuming $T_{\rm e} = 10^{5}$ K.  Inverse Compton cooling is taken into account, using the $r/R$-band light curve \citep{jg23} as a proxy for the bolometric light curve of \sn. While the model here adopts the self-similar solution for the shock-wave dynamics \citep{Chevalier98}, it reproduces the results of the more detailed calculation by \citet{Matsuoka19}. 

Adopting our fiducial values of $\epsilon_{\rm e} = \epsilon_{\rm B} = 0.1$, we find that the 230 GHz upper limits for \sn\ rule out $\dot{M}\sim 5\times 10^{-6}-2\times 10^{-3}$ M$_\odot$ yr$^{-1}$, comparable to the model exclusion region shown in Figure~\ref{fig:mdot}.  A key point is that the model accounts for the deceleration of the outer ejecta, which is important in the case of the high mass loss rate regime (this means that we need to evaluate the constraints on mass loss rate at somewhat different velocities for the lower and upper bounds in Figure~\ref{fig:mdot}).  Additionally, we stress that the choice of (unconstrained) microphysical parameters ($\epsilon_{\rm e}$ and $\epsilon_{\rm B}$) are important, as demonstrated in the right panel of Figure~\ref{fig:lcmodel} where we adopt the smaller values inferred for SN\,2020oi \citep{Maeda2021}. In this case, the allowed range of the mass-loss rate is $\dot M\lesssim 5 \times 10^{-5}$ M$_\odot$ yr$^{-1}$ or $\gtrsim 2 \times 10^{-3} M_\odot$ yr$^{-1}$; interestingly, an intermediate mass-loss rate of $\sim 10^{-4}$ M$_\odot$ yr$^{-1}$ is marginally allowed. In any case, it is important to use additional information such as the inverse Compton cooling or multi-wavelength information to robustly constrain the mass-loss rate \citep[e.g.,][]{Maeda23}. 

In recent work, \citet{jg23} estimated a mass loss rate of $\dot{M}\sim 10^{-2}$ M$_\odot$ yr$^{-1}$ to a scale of $\sim 10^{15}$ cm by comparing the ionization features in early optical spectra (extending to about 14 days) with spectral models of CSM interaction. This result is marked in Figure~\ref{fig:mdot} and is in agreement with the results presented here.  On the other hand, \citet{Grefenstette23} used hard X-ray data at 4 and 11 days to estimate $\dot{M}\sim 3\times 10^{-4}$ M$_\odot$ yr$^{-1}$ on a similar radial scale (or $\sim 7\times 10^{-4}$ M$_\odot$ yr$^{-1}$ if using $v_w=115$ km s$^{-1}$; see Figure~\ref{fig:mdot}).  This value is in tension with our results for $T_e=10^5$ K and marginally for $T_e=10^4$ K. However, it is essential to note that each of these approaches to determining the CSM density makes key simplifying assumptions that may lead to disparate estimates of the CSM properties. This includes numerical factors (e.g., $\epsilon_e$, $\epsilon_B$, and $T_e$ in our case) and underlying geometrical assumptions such as spherical symmetry and a homogeneous CSM. For example, it is possible that a non-homogeneous and radially-confined CSM, which may also affect the ejecta velocity angular profile, would lead to X-ray and mm emission regions of different densities.  Indeed, an asymmetric and confined CSM on a radial scale of $\sim 10^{15}$ cm has been inferred from high resolution optical spectroscopy \citep{Smith23}.

\section{Conclusions} 
\label{sec:conc}

We have presented mm-band observations of the recently-discovered \sn, covering about 2.7 to 18.6 days post-explosion. The non-detections place an upper bound on the luminosity at 230 GHz of $\lesssim 8.6\times 10^{25}$ erg s$^{-1}$ Hz$^{-1}$ at 2.7 and 7.7 days, and $\lesssim 3.4\times 10^{25}$ erg s$^{-1}$ Hz$^{-1}$ at 18.6 days.  These limits are about a factor of 2 times lower than the mm-band emission detected from SN\,2011dh (IIb) and about an order of magnitude lower than the emission in SN\,1993J (IIb) and SN\,2018ivc (IIL).  Using these limits we place constraints on the proximate CSM of the progenitor of \sn, out to a scale of $\sim {\rm few}\times 10^{15}$ cm of $\dot{M}\gtrsim 10^{-2}$ M$_\odot$ yr$^{-1}$ (free-free absorption) or $\lesssim 10^{-6}$ M$_\odot$ yr$^{-1}$ (synchrotron self-absorption).  We note that these limits are in agreement with the inferences from early optical spectroscopy \citep{jg23,Smith23} but are in tension with the inferred mass loss based on early X-ray observations \citep{Grefenstette23}.  A likely scenario that may alleviate this tension is a non-homogeneous and confined dense CSM \citep{Smith23}, which can lead to the X-ray emission emerging from lower density regions while the mm-band emission is absorbed.

We anticipate that continued observations of \sn\ in the radio/mm and X-ray regimes over the coming weeks, months, and years, as well as joint modeling with the available optical photometry and spectroscopy, will eventually better delineate the complex and likely non-spherical and non-homogeneous CSM environment of the progenitor.  More broadly, as demonstrated here, early mm-band observations can provide constraints on the density and geometry of the CSM around SN progenitors on the same spatial scales as rapid optical spectroscopy and X-ray observations, providing an independent constraint with different modeling assumptions.  We plan to continue undertaking such observations for nearby CCSNe as part of {\it POETS}.

\acknowledgments
The Berger Time-Domain research group at Harvard is supported by the NSF and NASA. T.E.~is supported by NASA through the NASA Hubble Fellowship grant HST-HF2-51504.001-A awarded by the Space Telescope Science Institute, which is operated by the Association of Universities for Research in Astronomy, Inc., for NASA, under contract NAS5-26555.

The Submillimeter Array is a joint project between the Smithsonian Astrophysical Observatory and the Academia Sinica Institute of Astronomy and Astrophysics and is funded by the Smithsonian Institution and the Academia Sinica.

\facilities{SMA}.


\end{document}